\documentclass[twocolumn,showpacs,superscriptaddress,amsmath,amssymb]{revtex4-1}
\usepackage{graphicx}
\usepackage{dcolumn}
\usepackage{bm}
\usepackage{natbib}
\usepackage{color}
\usepackage[colorlinks,
            bookmarks=true,
            linkcolor=blue,
            urlcolor=blue,
            anchorcolor=black,
            citecolor=blue
            ]{hyperref}
\usepackage[all]{hypcap}

\usepackage{hypcap}

\begin{document}

\title{Collision Energy Dependence of Moments of Net-Kaon Multiplicity Distributions at RHIC}

\affiliation{AGH University of Science and Technology, FPACS, Cracow 30-059, Poland}
\affiliation{Argonne National Laboratory, Argonne, Illinois 60439}
\affiliation{Brookhaven National Laboratory, Upton, New York 11973}
\affiliation{University of California, Berkeley, California 94720}
\affiliation{University of California, Davis, California 95616}
\affiliation{University of California, Los Angeles, California 90095}
\affiliation{Central China Normal University, Wuhan, Hubei 430079}
\affiliation{University of Illinois at Chicago, Chicago, Illinois 60607}
\affiliation{Creighton University, Omaha, Nebraska 68178}
\affiliation{Czech Technical University in Prague, FNSPE, Prague, 115 19, Czech Republic}
\affiliation{Nuclear Physics Institute AS CR, 250 68 Prague, Czech Republic}
\affiliation{Frankfurt Institute for Advanced Studies FIAS, Frankfurt 60438, Germany}
\affiliation{Institute of Physics, Bhubaneswar 751005, India}
\affiliation{Indiana University, Bloomington, Indiana 47408}
\affiliation{Alikhanov Institute for Theoretical and Experimental Physics, Moscow 117218, Russia}
\affiliation{University of Jammu, Jammu 180001, India}
\affiliation{Joint Institute for Nuclear Research, Dubna, 141 980, Russia}
\affiliation{Kent State University, Kent, Ohio 44242}
\affiliation{University of Kentucky, Lexington, Kentucky 40506-0055}
\affiliation{Lamar University, Physics Department, Beaumont, Texas 77710}
\affiliation{Institute of Modern Physics, Chinese Academy of Sciences, Lanzhou, Gansu 730000}
\affiliation{Lawrence Berkeley National Laboratory, Berkeley, California 94720}
\affiliation{Lehigh University, Bethlehem, Pennsylvania 18015}
\affiliation{Max-Planck-Institut fur Physik, Munich 80805, Germany}
\affiliation{Michigan State University, East Lansing, Michigan 48824}
\affiliation{National Research Nuclear University MEPhI, Moscow 115409, Russia}
\affiliation{National Institute of Science Education and Research, Bhubaneswar 751005, India}
\affiliation{National Cheng Kung University, Tainan 70101 }
\affiliation{Ohio State University, Columbus, Ohio 43210}
\affiliation{Institute of Nuclear Physics PAN, Cracow 31-342, Poland}
\affiliation{Panjab University, Chandigarh 160014, India}
\affiliation{Pennsylvania State University, University Park, Pennsylvania 16802}
\affiliation{Institute of High Energy Physics, Protvino 142281, Russia}
\affiliation{Purdue University, West Lafayette, Indiana 47907}
\affiliation{Pusan National University, Pusan 46241, Korea}
\affiliation{Rice University, Houston, Texas 77251}
\affiliation{University of Science and Technology of China, Hefei, Anhui 230026}
\affiliation{Shandong University, Jinan, Shandong 250100}
\affiliation{Shanghai Institute of Applied Physics, Chinese Academy of Sciences, Shanghai 201800}
\affiliation{State University of New York, Stony Brook, New York 11794}
\affiliation{Temple University, Philadelphia, Pennsylvania 19122}
\affiliation{Texas A\&M University, College Station, Texas 77843}
\affiliation{University of Texas, Austin, Texas 78712}
\affiliation{University of Houston, Houston, Texas 77204}
\affiliation{Tsinghua University, Beijing 100084}
\affiliation{University of Tsukuba, Tsukuba, Ibaraki, Japan,305-8571}
\affiliation{Southern Connecticut State University, New Haven, Connecticut 06515}
\affiliation{University of California, Riverside, California 92521}
\affiliation{United States Naval Academy, Annapolis, Maryland 21402}
\affiliation{Valparaiso University, Valparaiso, Indiana 46383}
\affiliation{Variable Energy Cyclotron Centre, Kolkata 700064, India}
\affiliation{Warsaw University of Technology, Warsaw 00-661, Poland}
\affiliation{Wayne State University, Detroit, Michigan 48201}
\affiliation{World Laboratory for Cosmology and Particle Physics (WLCAPP), Cairo 11571, Egypt}
\affiliation{Yale University, New Haven, Connecticut 06520}
\author{L.~Adamczyk}\affiliation{AGH University of Science and Technology, FPACS, Cracow 30-059, Poland}
\author{J.~R.~Adams}\affiliation{Ohio State University, Columbus, Ohio 43210}
\author{J.~K.~Adkins}\affiliation{University of Kentucky, Lexington, Kentucky 40506-0055}
\author{G.~Agakishiev}\affiliation{Joint Institute for Nuclear Research, Dubna, 141 980, Russia}
\author{M.~M.~Aggarwal}\affiliation{Panjab University, Chandigarh 160014, India}
\author{Z.~Ahammed}\affiliation{Variable Energy Cyclotron Centre, Kolkata 700064, India}
\author{N.~N.~Ajitanand}\affiliation{State University of New York, Stony Brook, New York 11794}
\author{I.~Alekseev}\affiliation{Alikhanov Institute for Theoretical and Experimental Physics, Moscow 117218, Russia}\affiliation{National Research Nuclear University MEPhI, Moscow 115409, Russia}
\author{D.~M.~Anderson}\affiliation{Texas A\&M University, College Station, Texas 77843}
\author{R.~Aoyama}\affiliation{University of Tsukuba, Tsukuba, Ibaraki, Japan,305-8571}
% ...
\collaboration{STAR Collaboration}\noaffiliation

\begin{abstract}
Fluctuations of conserved quantities such as baryon number, charge, and strangeness are sensitive to the correlation length of the hot and dense matter created in relativistic heavy-ion collisions and can be used to search for the QCD critical point. 
We report the first measurements of the moments of net-kaon multiplicity distributions in Au+Au collisions at $\sqrt{s_{\rm NN}}$ = 7.7, 11.5, 14.5, 19.6, 27, 39, 62.4, and 200 GeV. 
The collision centrality and energy dependence of the mean ($M$), variance ($\sigma^2$), skewness ($S$), and kurtosis ($\kappa$) for net-kaon multiplicity distributions as well as the ratio $\sigma^2/M$ and the products $S\sigma$ and $\kappa\sigma^2$ are presented. 
Comparisons are made with Poisson and negative binomial baseline calculations as well as with UrQMD, a transport model (UrQMD) that does not include effects from the QCD critical point. 
Within current uncertainties, the net-kaon cumulant ratios appear to be monotonic as a function of collision energy.\end{abstract}


\maketitle
\section{Introduction}
One primary goal of high energy heavy-ion collisions is to explore the phase structure of strongly interacting hot, dense nuclear matter. It can be displayed in the quantum chromodynamics (QCD) phase diagram, which is characterized by the temperature ($T$) and the baryon chemical potential ($\mu_{B}$). 
Lattice QCD calculations suggest that the phase transition between the hadronic phase and the quark-gluon plasma (QGP) phase at large $\mu_{B}$ and low $T$ is of the first order \cite{Ejiri:2008ir,Alford:1998ju}, while in the low $\mu_{B}$ and high $T$ region, the phase transition is a smooth crossover \cite{Aoki:2006gr}.
The end point of the first order phase boundary towards the crossover region is the so-called critical point~\cite{location,Gavai:2016eg}. 
Experimental search for the critical point is one of the central goals of the beam energy scan (BES) program at the Relativistic Heavy-Ion Collider (RHIC) facility at Brookhaven National Laboratory.

Fluctuations of conserved quantities, such as baryon number ($B$), charge ($Q$), and strangeness ($S$) are sensitive to the QCD phase transition and QCD critical point~\cite{Stephanov:2009fva,Stephanov:2011el,Gupta:2011dn}. 
Experimentally, one can measure the moments (mean ($M$), variance ($\sigma^2$), skewness ($S$), and kurtosis ($\kappa$)) of the event-by-event net-particle distributions (particle multiplicity minus antiparticle multiplicity), such as net-proton, net-kaon and net-charge multiplicity distributions in heavy-ion collisions. 
These moments are connected to the thermodynamic susceptibilities that can be computed in lattice QCD \cite{Gavai:2008er,Gavai:2016eg,Cheng:2009cx,Bazavov:2012ej,Bazavov:2012jj,Borsanyi:2013el,Alba:2014dn,KARSCH2016352} and in the hadron resonance gas model (HRG) \cite{Karsch:2011dh,Garg:2013jq,Fu:2013en,Nahrgang:2015gs}.
They are expected to be sensitive to the correlation length ($\xi$) of the hot and dense medium created in the heavy-ion collisions \cite{Stephanov:2009fva}. Non-monotonic variation of fluctuations in conserved quantities with the colliding beam energy is considered to be one of the characteristic signature of the QCD critical point.  

\begin{figure*}
\centering
\includegraphics[width=7in]{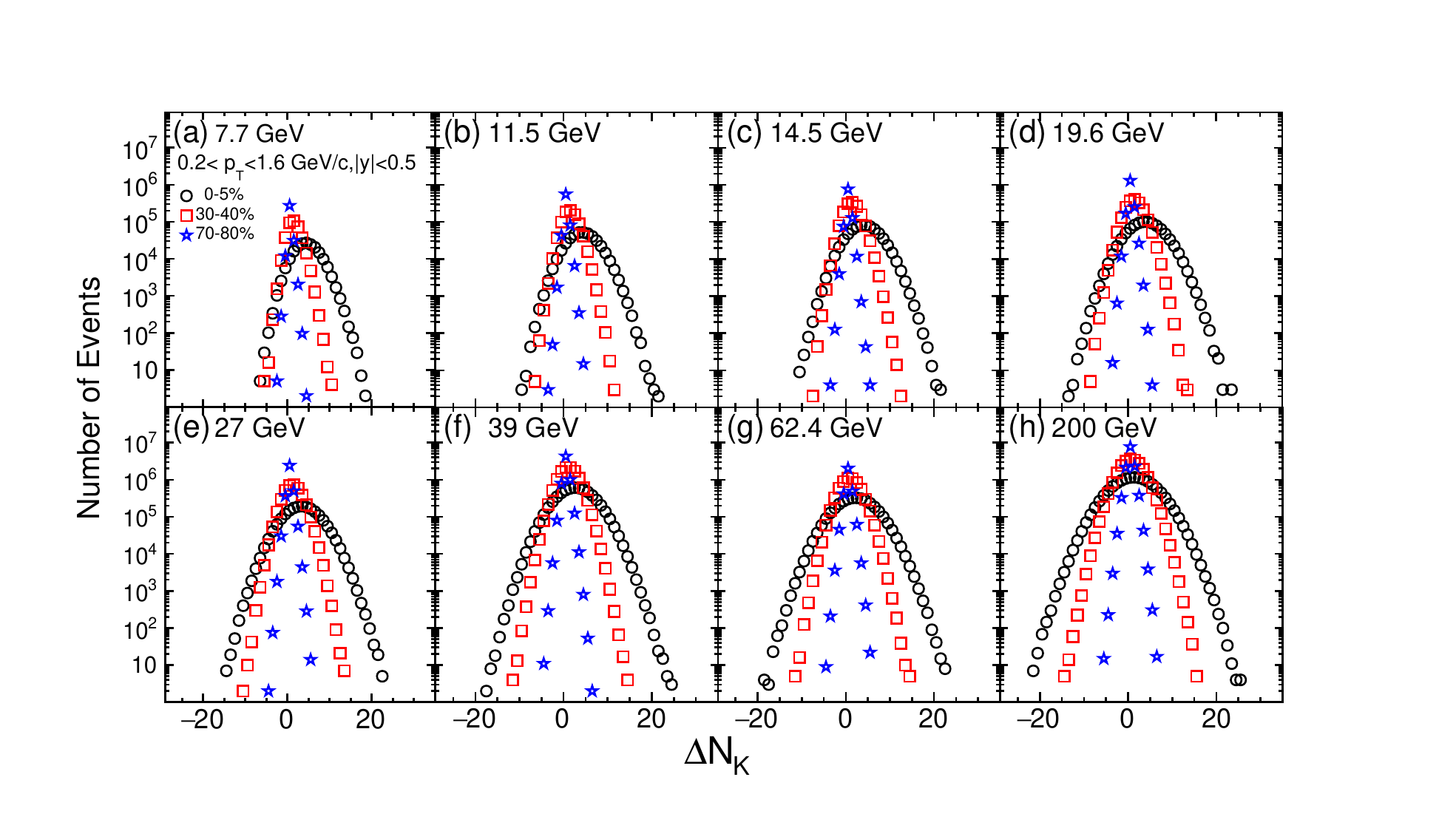}
\caption{(Color Online). Raw $\Delta{N_{K}}$ distributions in Au+Au collisions from $\sqrt{s_{\rm NN}}$ = 7.7 to 200 GeV for 0-5\%, 30-40\%, and 70-80\% collision centralities at midrapidity. 
The distributions are not corrected for the finite centrality bin width effect nor the reconstruction efficiency.}
\label{distribution}
\end{figure*}

The moments $\sigma^2$, $S$, and $\kappa$ have been shown to be related to powers of the correlation length as $\xi^2$, $\xi^{4.5}$ and $\xi^7$  \cite{Stephanov:2009fva}, respectively. 
The $n^{\rm th}$ order susceptibilities $\chi ^{(n)}$ are related to cumulant as ${\chi ^{(n)}} = C_n^{}/V{T^3}$ \cite{Gupta:2011dn}, where $V,T$ are the volume and temperature of the system, $C_{n}$ is the $n^{th}$ order cumulant of multiplicity distributions. 
The moment products $S\sigma$ and $\kappa\sigma^2$ and the ratio $\sigma^2/M$ are constructed to cancel the volume term. 
The moment products are related to the ratios of various orders of susceptibilities according to  $\kappa\sigma^2$=$\chi_{i}^{(4)}/\chi_{i}^{(2)}$ and $S\sigma$=$\chi_{i}^{(3)}/\chi_{i}^{(2)}$, where $i$ indicates the conserved quantity. 
Due to the sensitivity to the correlation length and the connection with the susceptibilities, one can use the moments of conserved-quantity distributions to aid in the search for the QCD critical point and the QCD phase transition \cite{Stephanov:2009fva,Karsch:2011dh,Friman:2014gt,Mukherjee:2015fm,NAHRGANG201683,Chen:2015dm,Chen:2016gv, Vovchenko:2015ib,PhysRevC.94.024918,Gupta:2011dn,Asakawa:2009be,Stephanov:2011el,Schaefer:2012eu,Adamczyk:2014ew,Adamczyk:2014kx,Gazdzicki:2017zrq}. 
In addition, the moments of net-particle fluctuations can be used to determine freeze-out points on the QCD phase diagram by comparing directly to first-principle lattice QCD calculations \cite{Bazavov:2012jj}. 
Specifically, by comparing the lattice QCD results to the measured $\sigma^{2}/M$ for net kaons, one can infer the hadronization temperature of strange quarks~\cite{1742-6596-779-1-012050}.

As a part of the BES, Au+Au collisions were run by RHIC with energies ranging from $\sqrt{s_{\rm NN}}$ = 200 GeV down to 7.7 GeV~\cite{Collaboration:2010uc,Abelev:2009bw,STAR_BES_note} corresponding  to $\mu_{B}$ from 20 to 420 MeV. 
In this paper, we report the first measurements for the moments of net-kaon multiplicity distributions in Au+Au collisions at $\sqrt{s_{\rm NN}}$ = 7.7, 11.5, 14.5, 19.6, 27, 39, 62.4, and 200 GeV. 
These results are compared with baseline calculations (Poisson and negative binomial) and the Ultrarelativistic Quantum Molecular Dynamics (UrQMD, version 2.3) model calculations \cite{Bleicher:1999fb}.

The manuscript is organized as follows.  In section II, we define the observables used in the analysis. 
In section III, we describe the STAR (Solenoidal Tracker At RHIC) experiment at BNL and the analysis techniques. 
In section IV, we present the experimental results for the moments of the net-kaon multiplicity distributions in Au+Au collisions at RHIC BES energies. A summary is given in section V.

\section{Observables}
Distributions can be characterized by the moments $M$, $\sigma^2$, $S$, and $\kappa$ as well as in terms of cumulants $C_1$, $C_2$, $C_3$, and $C_4$~\cite{Hald:2000ir}.

In the present analysis, we use $N$ to represent particle multiplicity in one event  and  $\Delta N_{K}$ ($N_{K^{+}} - N_{K^{-}}$) the net-kaon number. 
The average value over the entire event ensemble is denoted by $\langle N \rangle$. Then the deviation of $N$ from its mean value can be written as  $\delta N= N - \langle N \rangle$.
The various order cumulants of event-by-event distributions of $N$ are defined as:

     \begin{eqnarray}
     &&C_{1} = \langle N \rangle \\                                           
     &&C_{2} = \langle (\delta N)^2 \rangle\\                             
     &&C_{3} = \langle (\delta N)^3 \rangle \\                          
     &&C_{4} = \langle (\delta N)^4 \rangle-3 \langle (\delta N)^2 \rangle^2
     \end{eqnarray}
     
The moments can be written in terms of the cumulants as:

\begin{eqnarray}
M = C_{1}, \sigma^2 = C_{2}, S=\frac{C_{3}}{(C_{2})^\frac{3}{2}}, \kappa = \frac{C_{4}}{(C_{2})^2}
\end{eqnarray}

In addition, the products of moments $\kappa\sigma^2$ and $S\sigma$ can be expressed in terms of cumulant ratios:

\begin{eqnarray}
\label{eq6} \kappa\sigma^2 = \frac{C_{4}}{C_{2}}, S\sigma = \frac{C_{3}}{C_{2}}
\end{eqnarray}

\section{DATA ANALYSIS}

The results presented in this paper are based on the data taken at STAR \cite{Ackermann2003} for Au+Au collisions at $\sqrt{\it{s_{\rm NN}}}$ = 7.7, 11.5, 14.5, 19.6, 27, 39, 62.4 and 200 GeV. 
The 7.7, 11.5, 39, 62.4, and 200 GeV data were collected in the year 2010, the 19.6 and 27 GeV data were collected in the year 2011, and the 14.5 GeV data were collected in the year 2014. 

The STAR  detector has a large uniform acceptance at midrapidity ($|\eta|<1$) with excellent particle identification capabilities
, i.e., allowing to identify kaons from other charged particles for $0.2 < p_{T} < 1.6$ GeV/$c$. 
Energy loss ($dE/dx$) in the time projection chamber (TPC) \cite{Anderson:2003jk} and mass-squared ($m^{2}$) from the time-of-flight detector (TOF) \cite{Llope:2005fx} are used to identify $K^{+}$ and $K^{-}$. To utilize the energy loss measured in the TPC, a quantity $n\sigma_{X}$ is defined as:

\begin{eqnarray}
n\sigma_{X}=\frac{\ln[(dE/dx)_{\rm measured}/(dE/dx)_{\rm theory}]}{\sigma_{X}}
\end{eqnarray}
 where $(dE/dx)_{\rm measured}$ is the ionization energy loss from TPC, and $(dE/dx)_{\rm theory}$ is the Bethe-Bloch \cite{Bichsel:2006ka} expectation for the given particle type (e.g. $\pi, K, p$). $\sigma_{X}$ is the $dE/dx$ resolution of TPC. 
We select $K^{+}$ and $K^{-}$ particles by using a cut $|n\sigma_{Kaon}| < 2$ within transverse momentum range $0.2 < p_{T} < 1.6$ GeV/$c$ and rapidity $|y| < 0.5$. 
The TOF detector measures the time of flight $(t)$ of a particle from the primary vertex of the collision. Combined with the path length ($L$) measured in the TPC, one can directly calculate the velocity $(v)$ of the particles and their rest mass $(m)$ using:

\begin{eqnarray}
&&\beta = \frac{v}{c} = \frac{L}{ct} \\%
&&{m^2c^2} = {p^2}\left( {\frac{1}{{{\beta ^2}}} - 1} \right) = {p^2}\left( {\frac{{{c^2}{t^2}}}{{{L^2}}} - 1} \right)
\end{eqnarray}

In this analysis, we use mass-squared cut $0.15 < m^{2} < 0.4$ GeV$^{2}/c^{4}$ to select $K^{+}$ and $K^{-}$ within the $p_{T}$ range $0.4 < p_{T} < 1.6$ GeV/$c$ to get high purity of kaon sample (better than 99\%).  For the $p_{T}$ range $0.2 < p_{T} < 0.4$ GeV/$c$, we use only the TPC to identify $K^{+}$ and $K^{-}$. The kaon purity between 0.2 and 0.4 GeV/c is about 95\%, where only TPC is used.
 \begin{figure*}
\centerline{\includegraphics[width=7in]{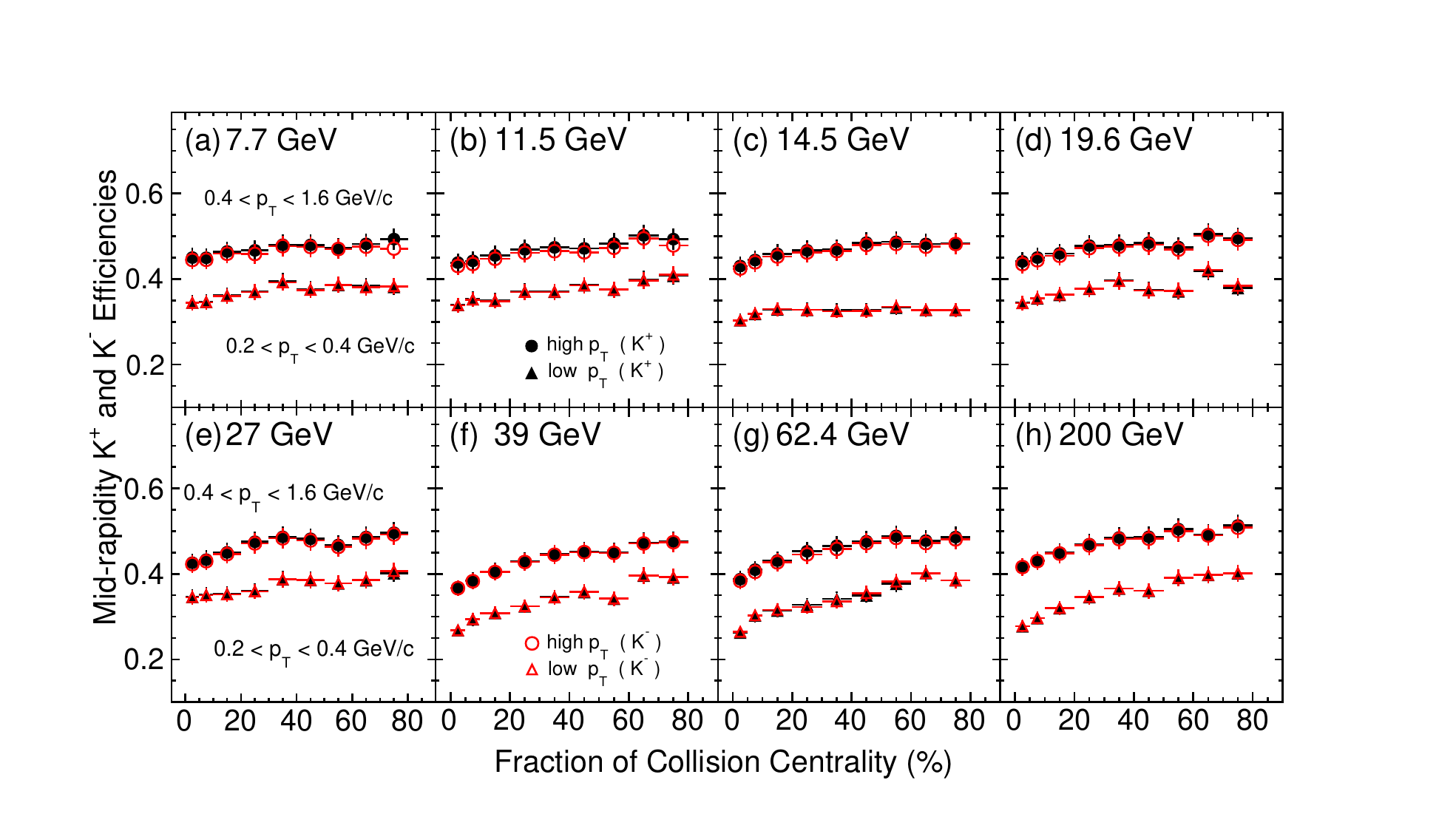}}
\caption{(Color Online). Collision centrality dependence of the $p_{T}$-averaged efficiencies in Au+Au collisions. For the lower $p_{T}$ range ($0.2<p_{T}<0.4$ GeV/$c$), only the TPC is used. For the higher $p_{T}$ range ($0.4<p_{T}<1.6$ GeV/$c$), both the TPC and TOF are used for particle identification (PID).}
\label{efficiency}
\end{figure*}

The collision centrality is determined using the efficiency-uncorrected charged particle multiplicity excluding  identified kaons within pseudorapidity $|\eta| < 1.0$ measured with the TPC.  This definition maximizes the number of particles used to determine the collision centrality and avoids self-correlations between the kaons used to calculate the moments and kaons in the reference multiplicity~\cite{Luo:2013cm}. Using the distribution of this reference multiplicity along with the Glauber model~\cite{Miller:2007ce} simulations, the collision centrality is determined. This reference multiplicity is similar in concept to the reference multiplicity used by STAR to study moments of net-proton distributions~\cite{Adamczyk:2014ew}, where the reference multiplicity was calculated using all charged particles within $|\eta| < 1.0$ excluding identified protons and antiprotons.  Using this definition, collision centrality bins of 0-5\%, 5-10\%, 10-20\%, 20-30\%, 30-40\%, 40-50\%, 50-60\%, 60-70\%, and 70-80\% of the multiplicity distributions were used with 0-5\% representing the most central collisions.

Figure~\ref{distribution} shows the raw event-by-event net-kaon multiplicity ($\Delta{N_{K}} = N_{K^{+}} - N_{K^{-}}$) distributions in Au+Au collisions at $\sqrt{s_{\rm NN}}$ = 7.7 to 200 GeV for three collision centralities, i.e. 0-5\%, 30-40\%, and 70-80\%. 
For the 0-5\% central collision, the peaks of the distributions are close to zero at high energies, and shift towards the positive direction as the energy decreases. 
This is because the pair production of $K^{+}$ and $K^{-}$ dominates at high energies while the production of $K^{+}$ is dominated by the associated production via reaction channel $NN \to N\Lambda K^{+}$ at lower energy~\cite{Margetis:2000sv}. Collision volume fluctuations within one finite centrality bin will lead to bin width effects, which can be corrected by applying the so-called centrality bin width correction~\cite{Luo:2013cm}. Those distributions are not corrected for the finite centrality bin width effect and also track reconstruction efficiency.  However, all the cumulants and their ratios presented in this paper are corrected for the finite centrality bin width effect and efficiency of $K^{+}$ and $K^{-}$.  

The moments and cumulants can be expressed in terms of factorial moments, which can be easily corrected for efficiency~\cite{Bzdak:2015ee,Luo:2015cxa}. The efficiency correction is done by assuming the response function of the efficiency is  a binomial probability distribution. Figure~\ref{efficiency} shows the collision centrality dependence of the $p_{T}$-averaged efficiencies of tracking and PID combined for two $p_{T}$ ranges. One can see that at the lower $p_{T}$ range ($0.2 < p_{T} < 0.4$ GeV/$c$), kaons have a lower efficiency compared with the higher $p_{T}$ range ($0.4 < p_{T} < 1.6$ GeV/$c$). The efficiencies increase monotonically with the centrality changing from most central (0 $\sim$ 5\%) to peripheral (70 $\sim$ 80\%). $K^{+}$ and $K^{-}$ have similar efficiencies. 

\begin{figure}
\centering
\includegraphics[width=3.7in]{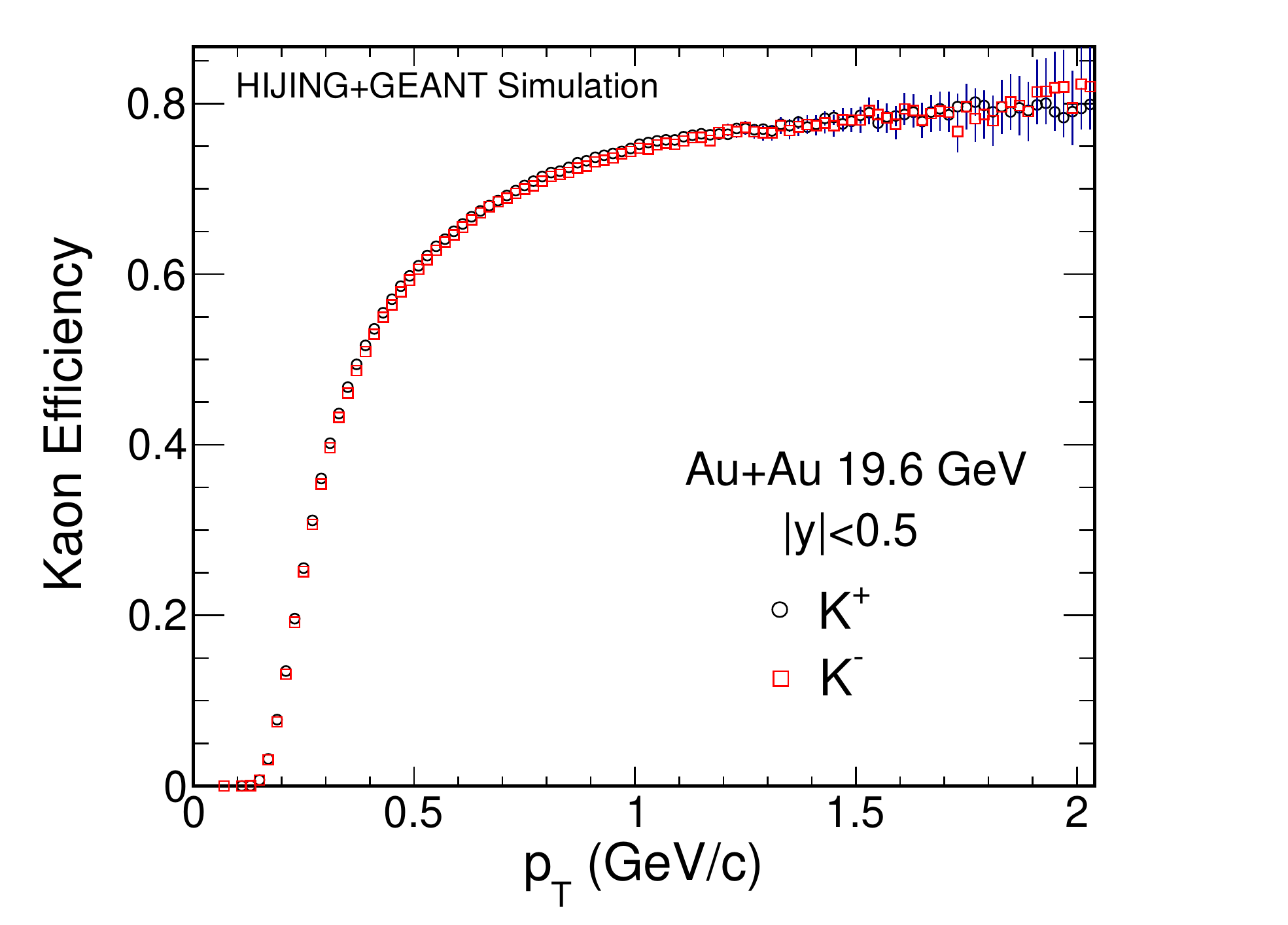}
\caption{(Color Online) Transverse momentum dependence of the efficiency (tracking+acceptance) for  
$K^+$ and $K^-$  in Au+Au collisions at 19.6 GeV from HIJING+GEANT simulation. }
\label{kaon_eff}
\end{figure}

\begin{figure}
\centering
\includegraphics[width=3.7in]{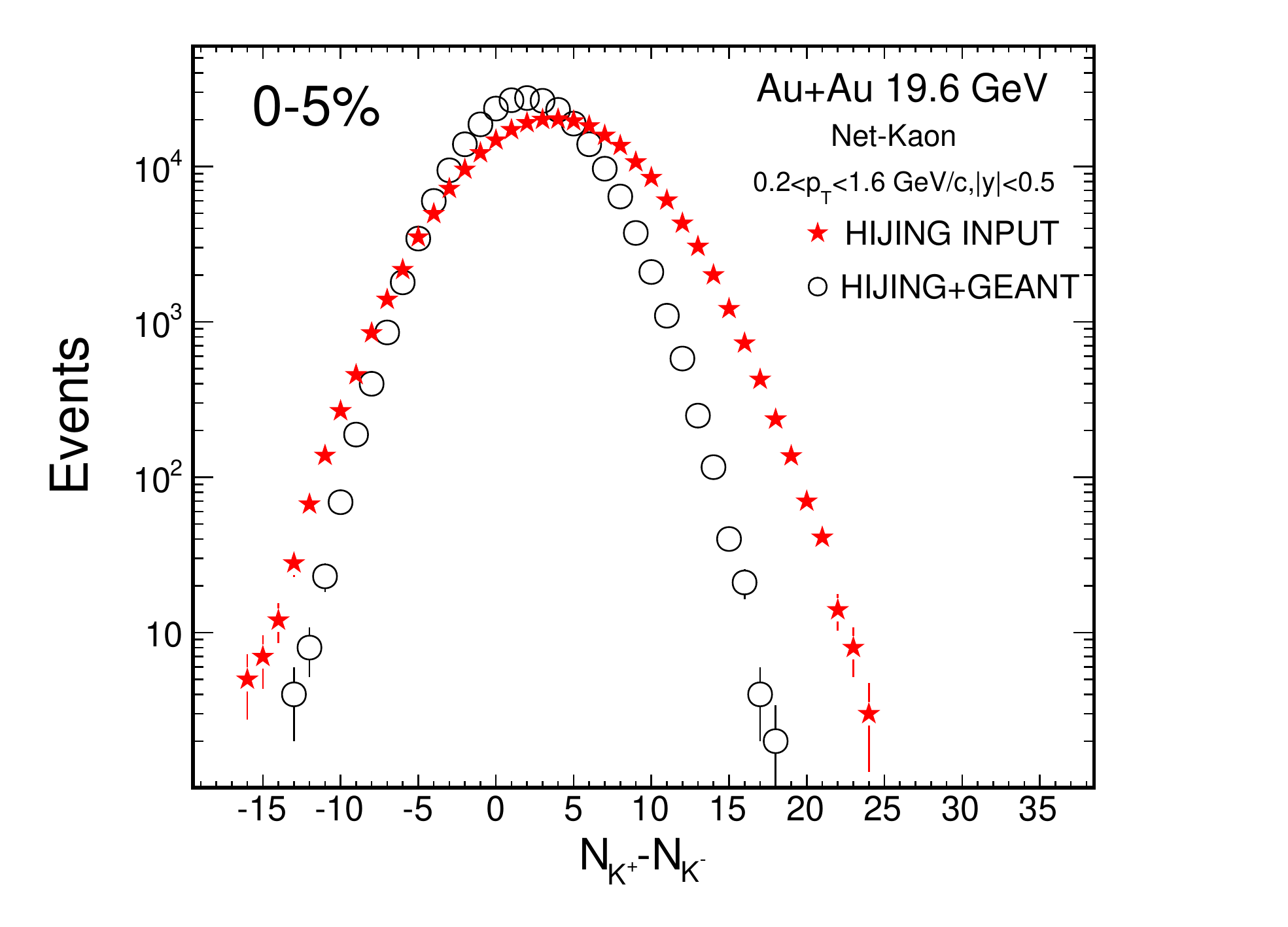}
\caption{(Color Online) Event-by-event net-kaon multiplicity distributions for 0-5\% Au+Au collisions at $\sqrt{s_{\rm NN}}$=19.6 GeV from HIJING input and HIJING+GEANT, respectively.  }
\label{netk_dis}
\end{figure}

\begin{figure*}
\centering
\includegraphics[width=7in]{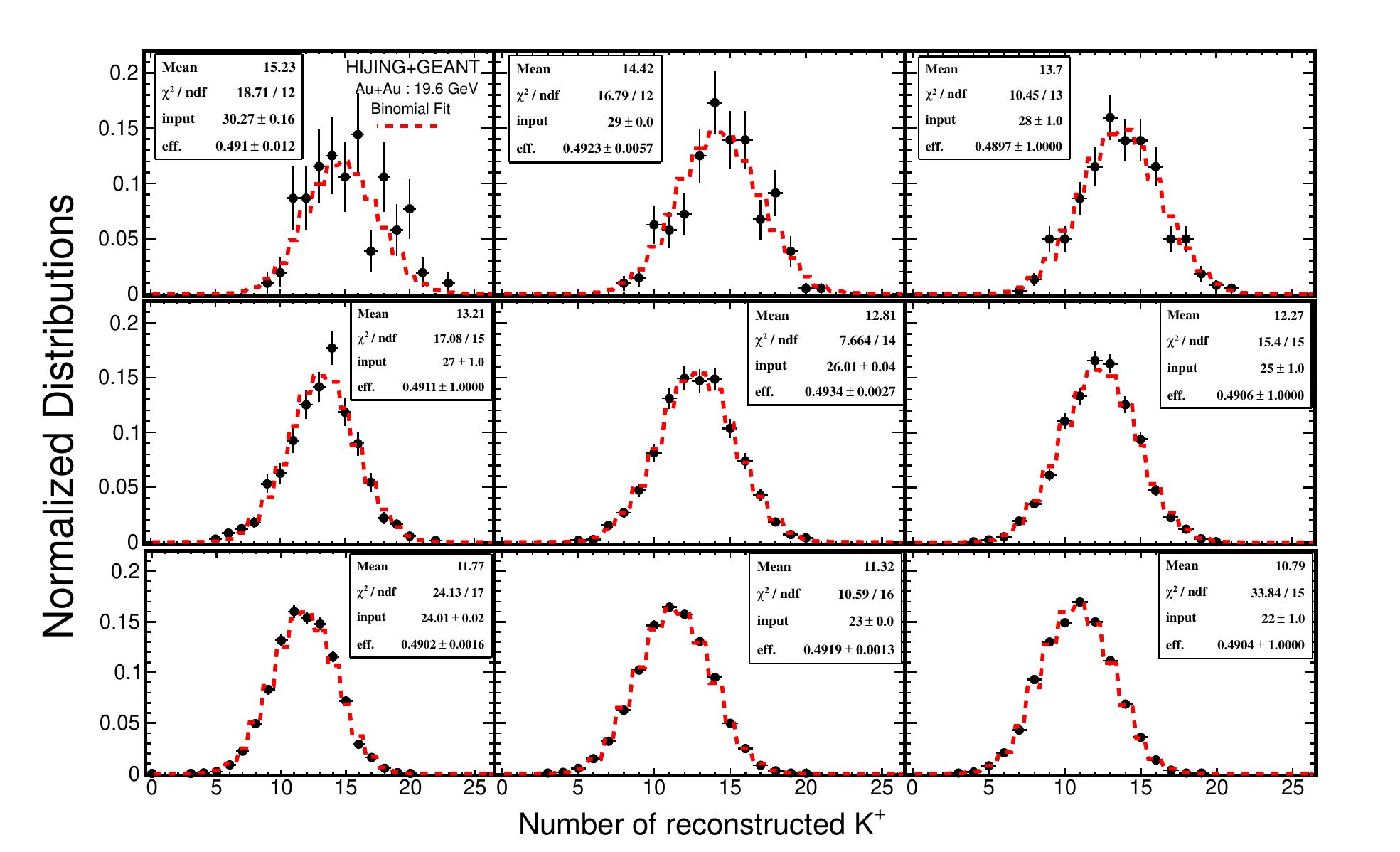}
\caption{(Color Online) Event-by-event reconstructed $K^{+}$ multiplicity distributions for Au+Au collisions at $\sqrt{s_{\rm NN}}$=19.6 GeV with various fixed input number of K+ (22-30) from HIJING, respectively. 
  }
\label{binomial_test}
\end{figure*}

\begin{figure*}
\centering
\includegraphics[width=7in]{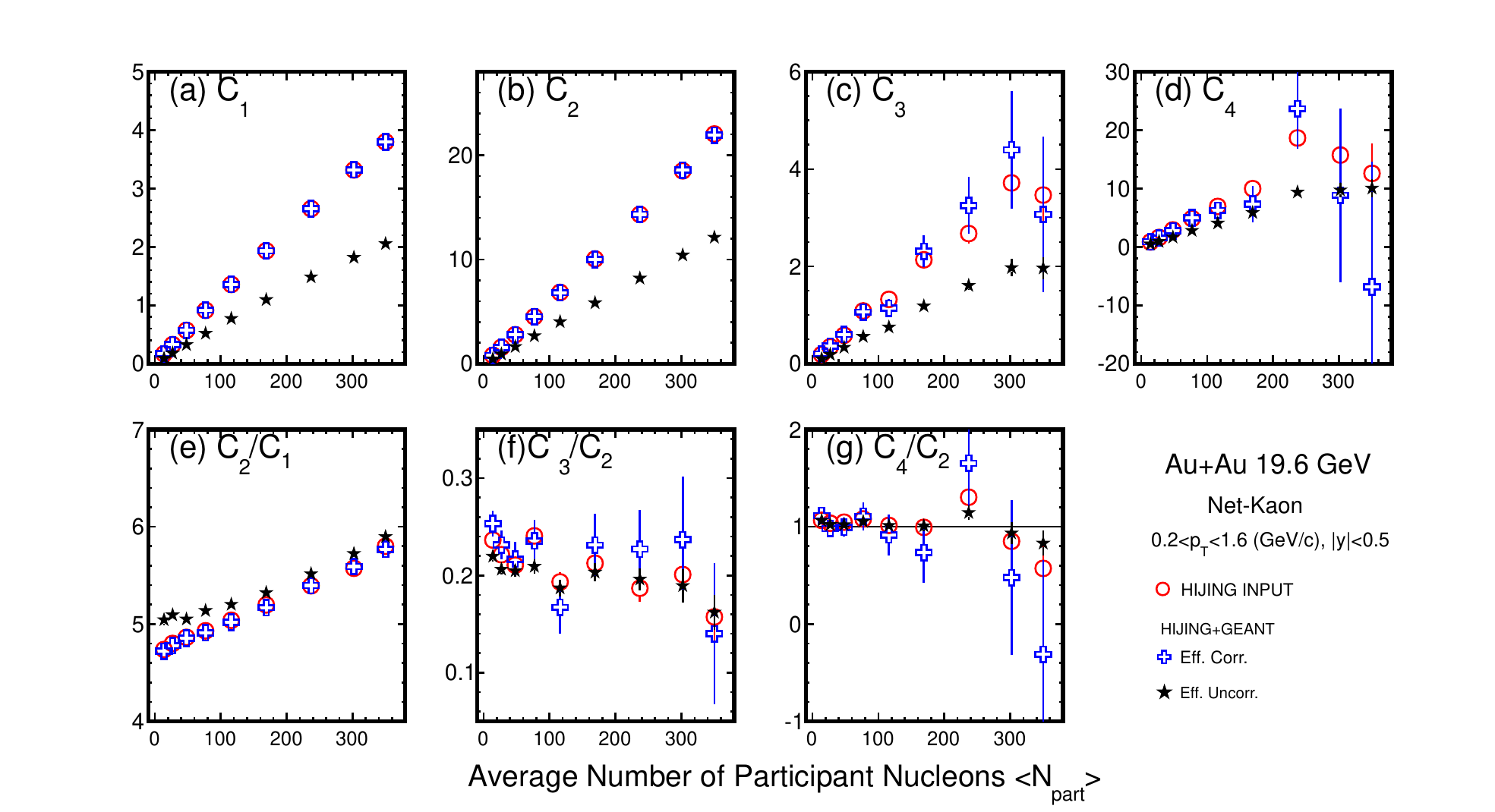}
\caption{(Color Online) Cumulants (1-4$^{th}$ order) of net-kaon distributions in Au+Au collisions at $\sqrt{s_{\rm NN}}$=19.6 GeV from HIJING+Geant simulations. The red circle represents the results from HIJING INPUT. The blue crosses and black stars are the results of efficiency corrected and uncorrected results after HIJING events passing through the GEANT simulation with realistic STAR detector environment. }
\label{cum_HJ}
\end{figure*}

\begin{figure*}
\hspace{-1cm}
\centerline{\includegraphics[width=7in]{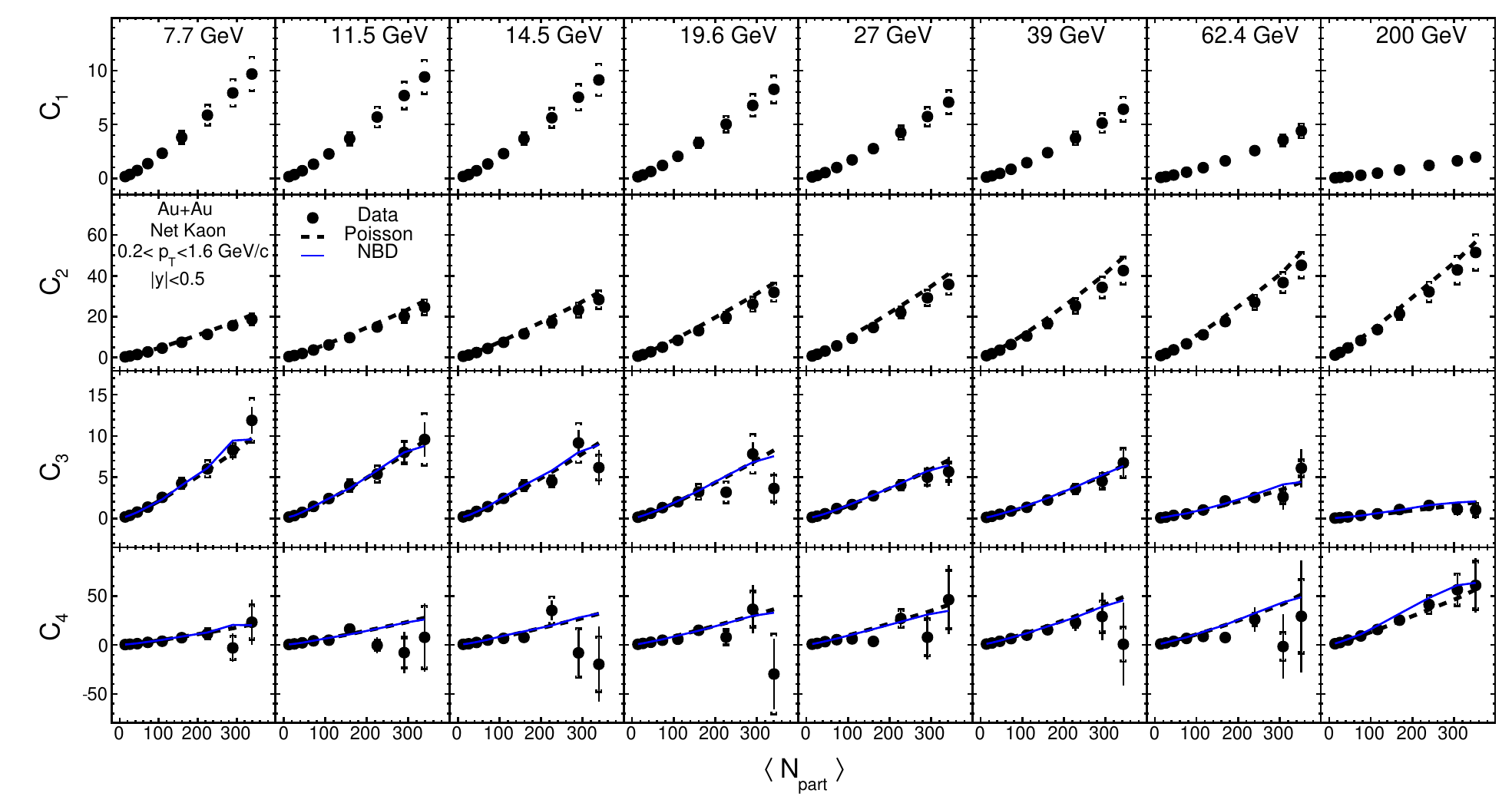}}
\caption{(Color Online). Collision centrality dependence of cumulants ($C_{1}$, $C_{2}$, $C_{3}$, and $C_{4}$) of $\Delta{N_{K}}$ distributions for Au+Au collisions at $\sqrt{s_{\rm NN}}$ = 7.7 - 200 GeV.  The error bars are statistical uncertainties and the caps represent systematic uncertainties. The Poisson and NBD expectations are shown as dashed and blue solid lines, respectively.}
\label{cumulants}
\end{figure*}

\begin{figure*}
\centerline{\includegraphics[width=7in]{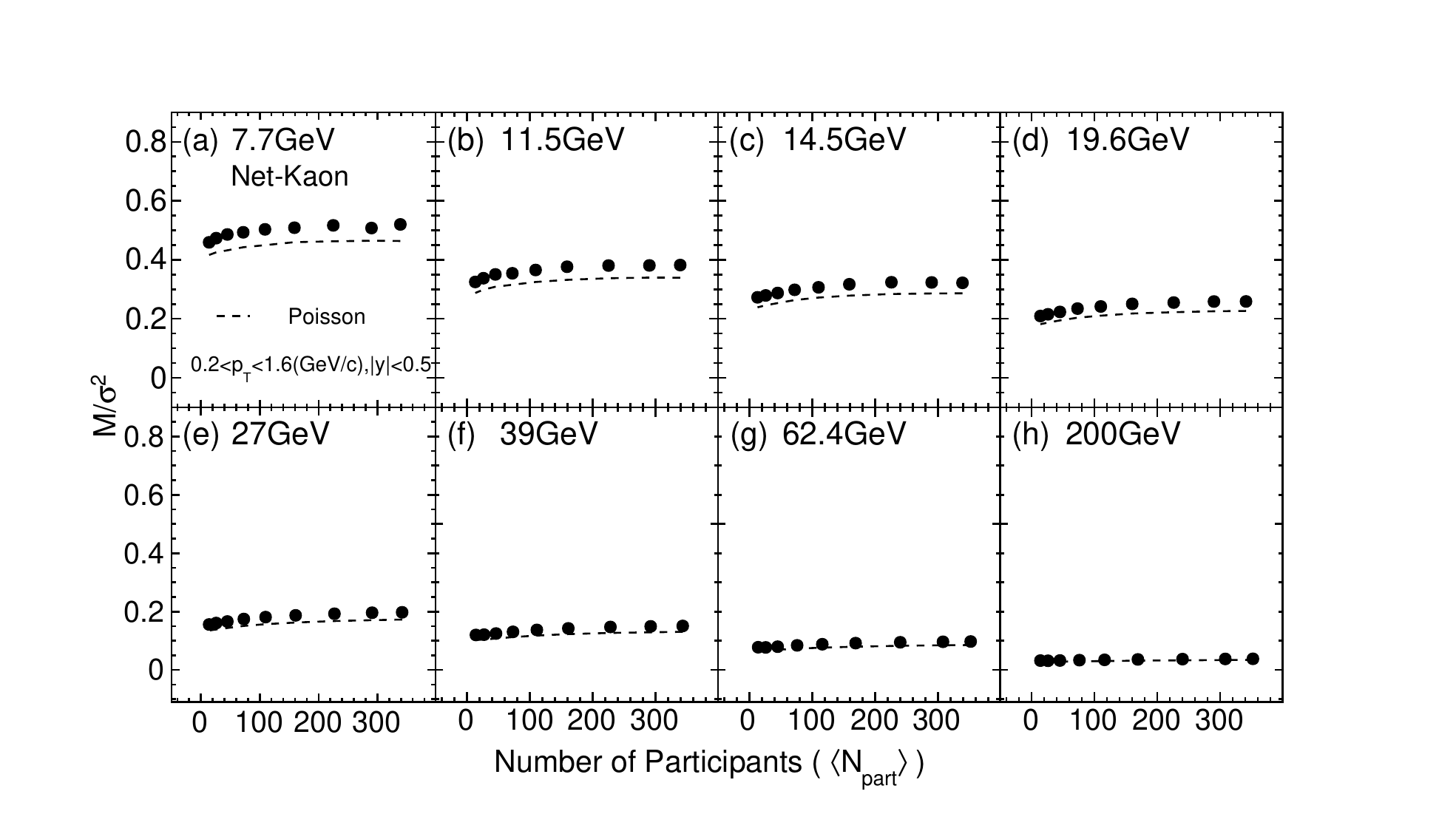}}
\caption{(Color Online).Collision centrality dependence of $M/\sigma^{2}$ for $\Delta{N_{K}}$ distributions in Au+Au collisions at $\sqrt{s_{\rm NN}}$ = 7.7 - 200 GeV. The Poisson expectations are shown as dashed lines. }
\label{VM}
\end{figure*}

\begin{figure*}
\centerline{\includegraphics[width=7in]{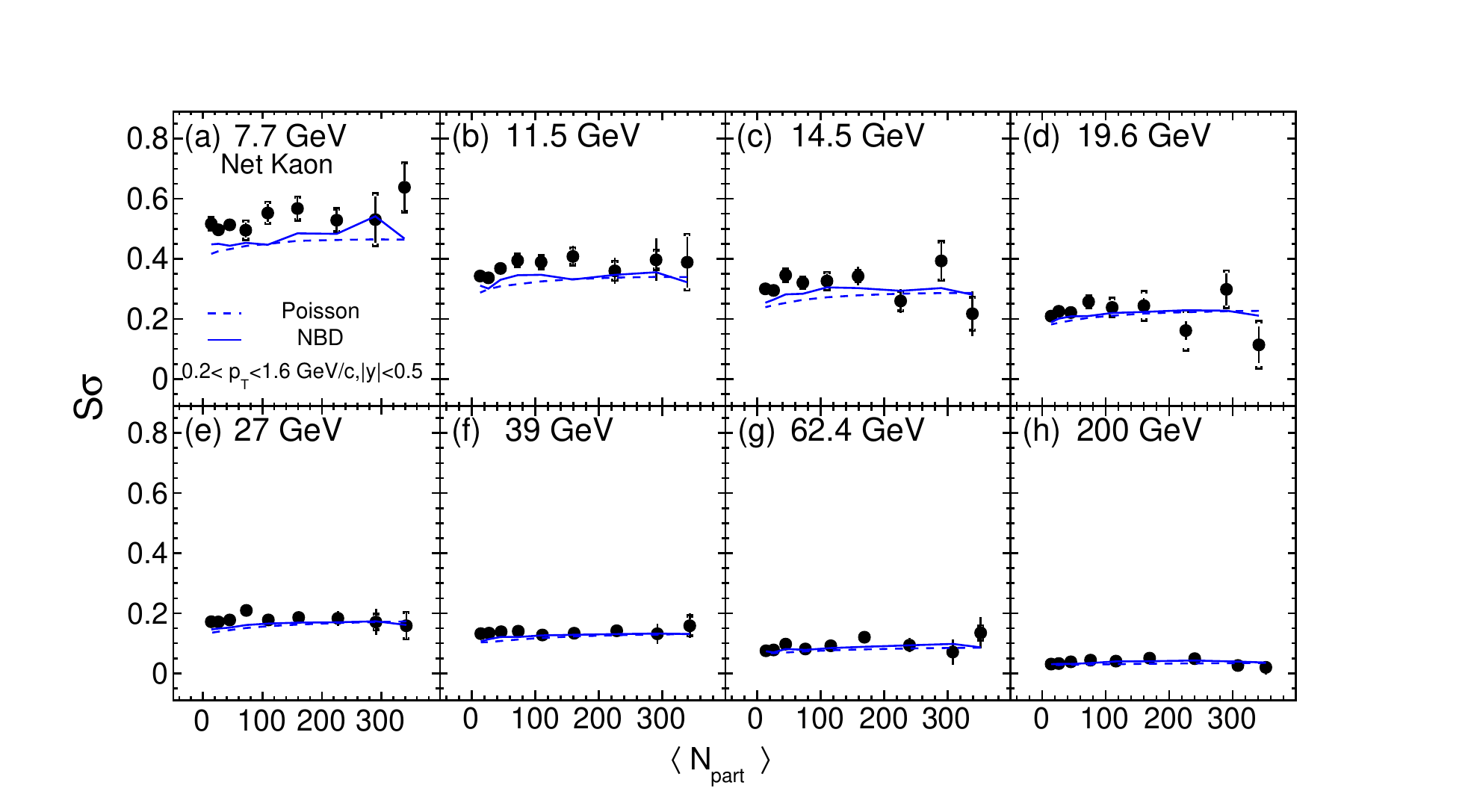}}
\caption{(Color Online). Collision centrality dependence of the $S\sigma$ for $\Delta{N_{K}}$ distributions from Au+Au collisions at $\sqrt{s_{\rm NN}}$ = 7.7 - 200 GeV. The error bars are statistical uncertainties and the caps represent systematic uncertainties. The Poisson (dashed line) and NBD (blue solid line) expectations are also shown. }
\label{SD}
\end{figure*}

\begin{figure*}
\centerline{\includegraphics[width=7in]{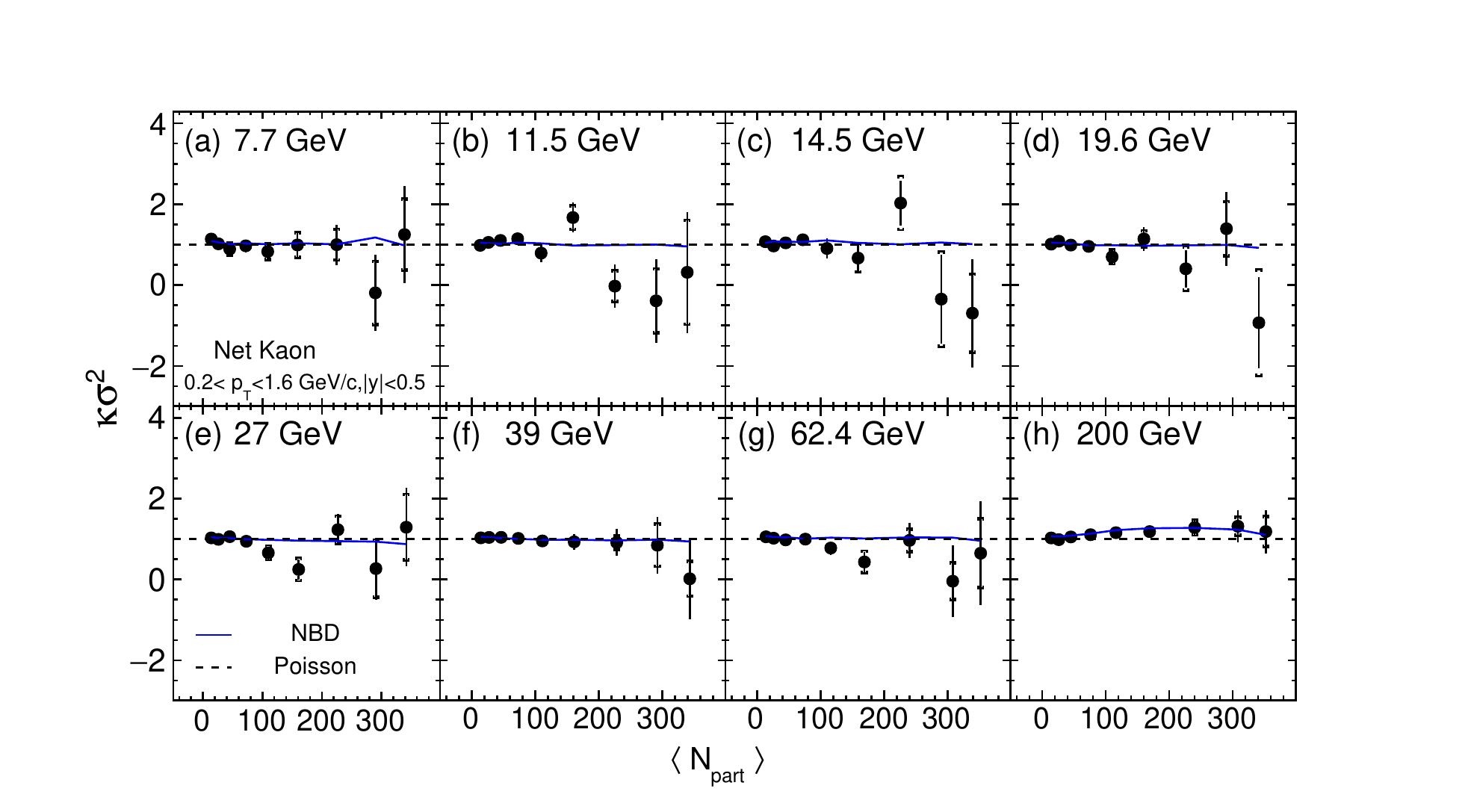}}
\caption{(Color Online). Collision centrality dependence of the $\kappa\sigma^{2}$ for $\Delta{N_{K}}$ distributions from Au+Au collisions at $\sqrt{s_{\rm NN}}$ = 7.7 - 200 GeV. The error bars are statistical uncertainties and the caps represent systematic uncertainties. The Poisson (dashed-line) and NBD (blue-solid-line) expectations are also shown.}
\label{KV}
\end{figure*}

\begin{figure}
\centerline{\includegraphics[width=3.5in]{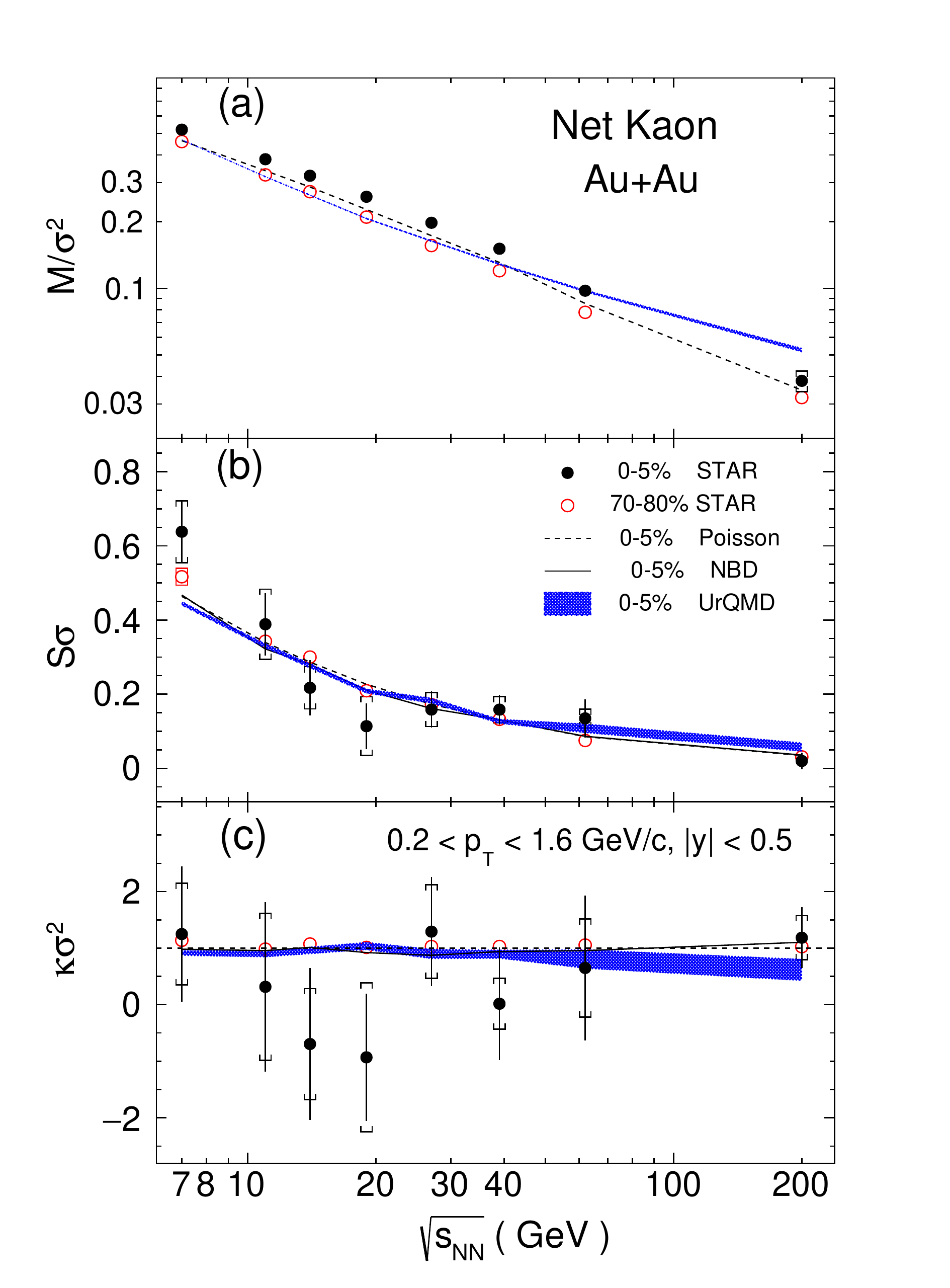}}
\caption{(Color Online). Collision energy dependence of the values of $M/\sigma^{2}$, $S\sigma$, $\kappa\sigma^{2}$ for $\Delta{N_{K}}$ multiplicity distributions from 0-5\% most central and 70-80\% peripheral collisions in Au+Au collisions at $\sqrt{s_{\rm NN}}$ = 7.7, 11.5, 14.5, 19.6, 27, 39, 62.4 and 200 GeV. The error bars are statistical uncertainties and the caps represent systematic uncertainties. The expectations from Poisson and NBD and the results of the UrQMD model calculations are all from the 0-5\% centrality. }
\label{energy}
\end{figure}
By calculating the covariance between the various order factorial moments, one can obtain the statistical uncertainties for the efficiency corrected moments based on the error propagation derived from the Delta theorem~\cite{Luo:2011tp,Luo:2013cm,Luo:2015cxa}. The statistical uncertainties of various order cumulants and cumulant ratios strongly depend on the width ($\sigma$) of the measured multiplicity distributions as well as the efficiencies ($\varepsilon$). One can roughly estimate the statistical uncertainties of $S\sigma$ and $\kappa \sigma^{2}$ as $error(S\sigma) \propto \frac{\sigma }{{{\varepsilon ^{3/2}}}} $ and $error(\kappa \sigma^{2}) \propto \frac{\sigma^2 }{{{\varepsilon ^{2}}}}$. That explains why we observe larger statistical uncertainties for central than peripheral collisions, as on the width of the net-kaon distributions grows from peripheral to central. Furthermore, due to the smaller detection efficiency of kaons than the protons, we observe larger statistical uncertainties of cumulants and cumulant ratios than those of the net-proton fluctuations~\cite{Adamczyk:2014ew}. 
Systematic uncertainties are estimated by varying the following track quality cuts: distance of closest approach, the number of fit points used in track reconstruction, the $dE/dx$ selection criteria for identification, and additional 5\% uncertainties in the reconstruction efficiency. The typical systematic uncertainties are of the order of 15\% for $C_1$ and $C_2$, 21\% for $C_3$, and 65\% for $C_4$. The statistical and systematic (caps) errors are presented separately in the figures.

\section{HIJING+GEANT Simulations}
To evaluate our methods of data analysis, we have done a standard STAR GEANT simulation with realistic detector environment and input from HIJING model. We generated 4.8 million mini-bias HIJING events for Au+Au collisions at $\sqrt{s_\mathrm{NN}}$=19.6 GeV and passed the particles from HIJING events through STAR detector simulated by using GEANT framework. The track reconstruction in the simulation is done with the same tracking algorithm as used in the STAR experiment.  We also applied the same track selection criteria, kinematic cuts ($0.2<p_T<1.6$ GeV, $|y|<0.5$)  for the reconstructed kaons ($K^+$ and $K^-$) as we used in the real data analysis. To avoid auto-correlation, the multiplicities of charged protons and pions within pseudo-rapidity range $|\eta|<1$ are used to define the collision centralities. The centrality bin width correction is also applied to suppress the volume fluctuations within wide centrality bins. As the charged particle ionization energy loss $dE/dx$ in the TPC gas is not simulated with sufficient details in the current STAR simulations and the time-of-flight (TOF) detectors are not implemented in the STAR Geant, we thus didn't include dE/dx and TOF identification efficiencies in the simulations. The matching of reconstructed and simulated tracks is determined by requiring at least 10 shared common hit points. By doing the simulation, we can test the analysis methods as well as the detector response in terms of the tracking performance with the realistic tracking environment and algorithm.

Figure~\ref{kaon_eff} shows the transverse momentum ($p_T$) dependence of the efficiency (track reconstruction +acceptance) for $K^+$ and $K^-$ in HIJING+GEANT simulation for Au+Au collisions at $\sqrt{s_\mathrm{NN}}$=19.6 GeV. These efficiencies are obtained by taking the ratio between the kaon $p_T$ spectra from HIJING+GEANT and HIJING input,  via the formula:
\begin{eqnarray} \label{eq:eff}
\langle \varepsilon \rangle  = \frac{{\int {{{\left( {\frac{{dN}}{{d{p_T}}}} \right)}_{\mathrm{RC}}}d{p_T}} }}{{\int {{{\left( {\frac{{dN}}{{d{p_T}}}} \right)}_{\mathrm{MC}}}d{p_T}} }}
\end{eqnarray}
where the RC and MC represent the tracks reconstructed from HIJING+GEANT and from HIJING input, respectively. Those average efficiencies for $K^+$ and $K^-$ are calculated within $0.2<p_T<1.6$ GeV/c,$|y|<0.5$ at nine centralities in Au+Au collisions at $\sqrt{s_\mathrm{NN}}$=19.6 GeV and are with the similar ($\sim50\%-55\%$) values as the TPC tracking efficiencies obtained from STAR embedding simulation. 

Figure~\ref{netk_dis} shows the event-by-event net-kaon multiplicity distributions for 0-5\% Au+Au collisions at $\sqrt{s_{\rm NN}}$=19.6 GeV from HIJING input and HIJING+GEANT, respectively. Due to the reconstruction efficiency loss of the kaons, the mean values and the width of net-kaon distributions become smaller than that of HIJING input. Based on the simulations, we have done a binomial test for the response function of kaon efficiencies. By fixing the input number of $K^{+}$, we fitted the event-by-event number of reconstructed $K^{+}$ distributions with binomial distribution function (red dashed lines). The fitting results are shown in Fig.~\ref{binomial_test} and results in different panels represent different input number of $K^{+}$, which is varied from 30 to 22 and corresponds to the 0-10\% collision centrality.  Fig.~\ref{cum_HJ} shows the centrality dependence of the cumulants and cumulant ratios of net-kaon multiplicity distributions in  Au+Au collisions at $\sqrt{s_{\rm NN}}$=19.6 GeV from the HIJING+GEANT simulations. The red circles represent the results from HIJING input. The blue crosses and black stars represent the efficiency corrected and uncorrected results, respectively. 
The efficiency correction is done by using the analytical formula derived from factorial moment assuming binomial response function of the kaon efficiencies~\cite{Bzdak:2015ee,Luo:2015cxa}. With this simplified simulation approach, the efficiency corrected cumulant and cumulant ratios are consistent with the results from the original HIJING INPUT for net-kaons. The efficiency correction procedure tends to significantly increase the statistical errors. The STAR collaboration is currently exploring other functions including a beta-binomial distribution to describe the efficiencies. 
Unfolding methods are being evaluated as correction method for efficiencies including those from PID cuts and tracking. 
The PID effects from the TPC and the TOF are not currently simulated through GEANT. 
These improvements will be necessary for the beam energy scan phase II program where an order of magnitude increase in the data sample is expected.

\section{Results}

Figure~\ref{cumulants} shows the centrality dependence of cumulants ($C_{1} -C_{4}$) of net-kaon ($\Delta{N_{K}}$) multiplicity distributions in Au+Au collisions at $\sqrt{s_\mathrm{NN}}$=7.7-200 GeV. The collisions centralities are represented by the average number of participating nucleons ($\langle N_{\rm part} \rangle$), which are obtained by Glauber model simulation.
The efficiency corrections have been done using the values shown in Fig.~\ref{efficiency}. 
In general, the various order cumulants show a nearly linear variation with $\langle N_{\rm part} \rangle$, which can be understood as the additivity property of the cumulants by increasing the volume of the system. This reflects the fact that the cumulants are extensive quantities that are proportional to the system volume. The decrease of the $C_{1}$ and $C_{3}$ values with increasing collision energy indicates that the ratio $K^{+}/K^{-}$ approaches unity for the higher collision energies. Figure~\ref{cumulants} also shows the Poisson and negative binomial distribution (NBD)~\cite{Luo:2014tga,Tarnowsky:2013dl} expectations. 
The Poisson baseline is constructed using the measured mean values of the multiplicity distributions of $K^{+}$ and $K^{-}$, while the NBD baseline is constructed using both means and variances. 
Assuming that the event-by-event multiplicities of $K^{+}$ and $K^{-}$ are independent random variables, the Poisson and NBD assumptions provide references for the moments of the net-kaon multiplicity distributions.
Within uncertainties, the measured cumulants values of $C_{3}$ and $C_{4}$ are consistent with both the Poisson and NBD baselines for most centralities.

The ratios between different order cumulants are taken to cancel the volume dependence. 
Figures~\ref{VM}, ~\ref{SD}, and ~\ref{KV} show the $\langle N_{\rm part} \rangle$ dependence of $\Delta{N_{K}}$ distributions for cumulant ratios $C_{1}/C_{2}$ (=$M/\sigma^{2}$), $C_{3}/C_{2}$ (=$S\sigma$), and $C_{4}/C_{2}$ (=$\kappa\sigma^{2}$), respectively. 
The values of $C_{1}/C_{2}$, shown in Fig.~\ref{VM}, systematically decrease with increasing collision energy for all centralities. 
The Poisson baseline for $C_{1}/C_{2}$ slightly underestimates the data,
indicating possible correlations between $K^{+}$ and $K^{-}$ production. 
For $C_{3}/C_{2}$ (=$S\sigma$) in Fig.~\ref{SD}, the Poisson and NBD expectations are observed to be lower than the measured $S\sigma$ values at low collision energies.
The measured values for $C_{4}/C_{2}$ (=$\kappa\sigma^{2}$) in Fig.~\ref{KV} are consistent with both the Poisson and NBD baselines within uncertainties.

The collision energy dependence of the cumulant ratios for $\Delta{N_{K}}$ distributions in Au+Au collisions are presented in Fig.~\ref{energy}. 
The results are shown in two collision centrality bins, one corresponding to most central (0-5\%) and the other to peripheral (70-80\%) collisions. 
Expectations from the Poisson and NBD baselines are derived for central (0-5\%) collisions. 
The values of $M/\sigma^{2}$ decrease as the collision energy increases, and are larger for central collisions compared with the peripheral collisions. 
For most central collisions, the Poisson baseline for $C_{1}/C_{2}$ slightly underestimates the data. 
Within uncertainties, the values of $S\sigma$ and $\kappa\sigma^{2}$ are consistent with both the Poisson and NBD baselines in central collisions. 
The blue bands give the results from the UrQMD model calculations for central (0-5\%) Au+Au collisions. 
The width of the bands represents the statistical uncertainties. 
The UrQMD calculations for $S\sigma$, and $\kappa\sigma^{2}$ are consistent with the measured values within uncertainties~\cite{Xu:2016jy,PhysRevC.96.014909}.  
A QCD based model calculation suggests that, due to heavy mass of the strange-quark, the sensitivity of the net-kaon ($\Delta N_K$) fluctuations is less than that of the net-proton ($\Delta N_p$)~\cite{Fan:2016ovc}. A much high statistics dataset is needed for the search of the QCD critical point with strangeness.

\section{SUMMARY}
In heavy-ion collisions, fluctuations of conserved quantities, such as net-baryon, net-charge and net-strangeness numbers, are sensitive observables to search for the QCD critical point. 
Near the QCD critical point, those fluctuations are expected to have similar energy dependence behavior. Experimentally, the STAR experiment has published the energy dependence of the net-proton (proxy for net-baryon)~\cite{Adamczyk:2014ew} and net-charge~\cite{Adamczyk:2014kx} fluctuations in Au+Au collisions from the first phase of the beam energy scan at RHIC. In this paper, we present the first measurements of the moments of net-kaon (proxy for net-strangeness) multiplicity distributions in Au+Au collisions from $\sqrt{s_{\rm NN}}$ = 7.7 to 200 GeV. The measured $M/\sigma^{2}$ values decrease monotonically with increasing collision energy. 
The Poisson baseline for $C_{1}/C_{2}$ slightly underestimates the data. 
No significant collision centrality dependence is observed for both $S\sigma$ and $\kappa\sigma^{2}$ at all energies. 
For $C_{3}/C_{2}$ (=$S\sigma$), the Poisson and NBD expectations are lower than the measured $S\sigma$ values at low collision energies. 
The measured values for $C_{4}/C_{2}$ (=$\kappa\sigma^{2}$) are consistent with both the Poisson and NBD baselines within uncertainties.
UrQMD calculations for $S\sigma$ and $\kappa\sigma^{2}$ are consistent with data for the most central 0-5\% Au+Au collisions. Within current uncertainties, the net-kaon cumulant ratios appear to be monotonic as a function of collision energy. 
The moments of net-kaon multiplicity distributions presented here can be used to extract freeze-out conditions in heavy-ion collisions by comparing to Lattice QCD calculations. Future high statistics measurements with improved efficiency correction method will be made for fluctuation studies in the second phase of the RHIC Beam Energy Scan during 2019-2020.

\section*{Acknowledgments}
We thank the RHIC Operations Group and RCF at BNL, the NERSC Center at LBNL, and the Open Science Grid consortium for providing resources and support. This work was supported in part by the Office of Nuclear Physics within the U.S. DOE Office of Science, the U.S. National Science Foundation, the Ministry of Education and Science of the Russian Federation, National Natural Science Foundation of China, Chinese Academy of Science,  the Ministry of Education and Science of the Russian Federation, National Natural Science Foundation of China, Chinese Academy of Science, the Ministry of Science and Technology of China (973 Programme No. 2015CB856900) and the Chinese Ministry of Education, the National Research Foundation of Korea, GA and MSMT of the Czech Republic, Department of Atomic Energy and Department of Science and Technology of the Government of India; the National Science Centre of Poland, National Research Foundation, the Ministry of Science, Education and Sports of the Republic of Croatia, RosAtom of Russia and German Bundesministerium fur Bildung, Wissenschaft, Forschung and Technologie (BMBF) and the Helmholtz Association. 

\bibliographystyle{mine}
\bibliography{mine}
\end{document}